\documentclass[final,5p,times,twocolumn,preprint]{elsarticle}
\usepackage{amssymb,amsmath}
\allowdisplaybreaks
\begin{document}
\begin{frontmatter}
\title{Nucleon matrix element of Weinberg's CP-odd gluon operator from the instanton vacuum}
\author{C. Weiss}
\address{Theory Center, Jefferson Lab, Newport News, VA 23606, USA}
\begin{abstract}
We calculate the nucleon matrix element of Weinberg's dimension-6 CP-odd gluon operator
$f^{abc} (\tilde F_{\mu\nu})^a (F^{\mu\rho})^b (F^{\nu}_{\;\;\rho})^c$
in the instanton vacuum. In leading order of the instanton packing fraction,
the dimension-6 operator is effectively proportional to the topological charge density
$(\tilde F_{\mu\nu})^a (F^{\mu\nu})^a$, whose nucleon matrix element
is given by the flavor-singlet axial charge and constrained by the $U(1)_A$ anomaly.
The nucleon matrix element of the dimension-6 operator is obtained substantially larger
than in other estimates, because of the strong localization of the nonperturbative gluon fields
in the instanton vacuum. We argue that the neutron electric dipole moment induced by the
dimension-6 operator is nevertheless of conventional size.
\end{abstract}
\end{frontmatter}
\section{Introduction}
The gluonic structure of the nucleon has become a topic of great interest in nuclear physics.
It is expressed in matrix elements of gauge-invariant composite QCD operators
between nucleon states and provides insight into the emergence of hadrons from color
fields. Gluonic operators arise in the factorization of deep-inelastic scattering (DIS)
processes and in the interaction of heavy quark systems with nucleons.
Gluonic operators also appear as the result of processes at the electroweak scale and beyond,
when the short-distance degrees of freedom are integrated out and represented by QCD operators,
which are then transported to the hadronic scale by the renormalization group equation.
In the context of CP-violation, one gluonic operator of particular interest is the
dimension-6 spin-0 CP-odd Weinberg operator \cite{Weinberg:1989dx}
\begin{align}
\mathcal{O}_{\rm P6} =
(g^3 / 16\pi^2) \, 
f^{abc} (\tilde G_{\mu\nu})^a (G^{\mu\rho})^b (G^{\nu}_{\;\;\rho})^c ,
\label{O6_def_G}
\end{align}
where $g$ is the QCD coupling, $f^{abc}$ are the $\textit{SU}(3)$ structure constants,
$(G_{\mu\nu})^a$ is the QCD field strength tensor and $(\tilde G_{\mu\nu})^a
\equiv (1/2) \epsilon_{\mu\nu\rho\sigma} (G^{\rho\sigma})^a$ its dual
(our conventions are given below; our definition of the operator
follows Ref.~\cite{Bigi:1990kz}). The operator Eq.~(\ref{O6_def_G}) appears in a scenario of
CP-violation that avoids the strong CP problem \cite{Peccei:2006as} and can be connected
to a nonzero electric dipole moment (EDM) of the nucleon \cite{Weinberg:1989dx,Bigi:1990kz},
whose experimental study is the object of extensive efforts \cite{Chupp:2017rkp}.
Because the operator cannot be measured directly in other processes,
the program relies on theoretical estimates of the hadronic matrix elements.
The operator Eq.~(\ref{O6_def_G}) also appears in the
$1/m_Q$ expansion of heavy quark contributions to hadron
structure \cite{Franz:1998hw,Polyakov:1998rb,Franz:2000ee,Polyakov:2015foa}.

Estimating hadronic matrix elements of gluonic operators such as Eq.~(\ref{O6_def_G})
is a challenging problem. Lattice QCD calculations are limited by the fact
that higher-dimensional operators mix with lower-dimensional ones of
the same quantum numbers with power-divergent coefficients, requiring accurate
non-perturbative treatment of the operator mixing \cite{Capitani:2002mp}.
Estimates using effective
models of hadron structure such as the quark model \cite{Yamanaka:2020kjo}
are uncertain because these models usually do not specify how the effective degrees
of freedom match with the non-perturbative gluon fields of QCD.
Methods based on QCD vacuum structure,
such as the estimate of Ref.~\cite{Bigi:1990kz} or the QCD sum rule calculations
of Refs.~\cite{Demir:2002gg,Haisch:2019bml}, can trace the connection between
QCD fields and hadronic structure and are appropriate to the problem,
but face difficulties in implementing the strong correlations affecting
matrix elements of higher-dimensional operators.
The approach of Refs.~\cite{Hatta:2020ltd,Hatta:2020riw} connects 
Eq.~(\ref{O6_def_G}) with operators appearing in higher-twist corrections
to the nucleon spin structure functions, but relies on additional assumptions for
estimating the matrix element (see below).

The instanton-based description of the QCD vacuum is a framework that can provide
realistic estimates of hadronic matrix elements of gluon operators such as Eq.~(\ref{O6_def_G}).
Its basic assumption is that the non-perturbative gluon fields relevant for hadron structure
are localized fluctuations carrying topological charge, instantons and antiinstantons,
with an average size $\bar\rho \sim 0.3$ fm much smaller than their average distance
$\bar R \sim$ 1 fm in 4-dimensional Euclidean spacetime \cite{Shuryak:1981ff,Shuryak:1982dp,Diakonov:1983hh}.
This is supported by a large body of empirical evidence from Euclidean correlation
functions and hadron structure, and by lattice QCD studies of field configurations using
cooling and other techniques; see Refs.~\cite{Schafer:1996wv,Diakonov:2002fq} for reviews.
In particular, instantons cause the spontaneous breaking of chiral symmetry,
through the fermionic zero modes associated with the topological charge,
which explains their unique role in hadron structure \cite{Diakonov:1985eg}.
Correlation functions and hadronic matrix elements
can be evaluated in an expansion
in the instanton packing fraction in the vacuum, $\bar\rho/\bar R$ (diluteness parameter),
which provides both a formal calculational scheme and an intuitive physical picture.

A method for evaluating gluonic operators in the instanton vacuum was formulated in
Ref.~\cite{Diakonov:1995qy} and used to calculate nucleon matrix elements of operators
such as the topological charge \cite{Diakonov:1995qy}, higher-twist operators appearing
in DIS \cite{Balla:1997hf,Dressler:1999zi,Lee:2001ug},
or the QCD energy-momentum tensor \cite{Polyakov:2018exb},
as well as higher-dimensional vacuum condensates \cite{Polyakov:1996kh,Polyakov:1998ip}.
The normalization scale of the operators in these calculations is set
by the average instanton size, $\mu \sim \bar\rho^{-1} \approx 0.6$ GeV, which
defines the boundary between perturbative and nonperturbative modes in
the instanton vacuum.
The expansion in the packing fraction ensures that the method preserves general operator relations
of QCD such as the low-energy theorems for the action density
$(G_{\mu\nu})^a (G^{\mu\nu})^a$ from the scale anomaly \cite{Diakonov:1995qy}
(see Ref.~\cite{Zahed:2021fxk} for a recent review),
the relation between the topological charge density
$g^2 (\tilde G_{\mu\nu})^a (G^{\mu\nu})^a$ and the flavor-singlet axial current
from the $U(1)_A$ anomaly \cite{Diakonov:1995qy},
and relations between higher-dimensional operators resulting from
the QCD equations of motion \cite{Balla:1997hf}. These properties are critically important in estimating
matrix elements of higher-dimensional gluonic operators such as Eq.~(\ref{O6_def_G}).

In this note we evaluate the nucleon matrix element of the gluonic operator Eq.~(\ref{O6_def_G})
in the instanton vacuum. In leading order of the packing fraction
(in the field of a single instanton), the operator Eq.~(\ref{O6_def_G}) is effectively proproportional
to the
topological charge density, whose matrix element is given by the nucleon flavor-singlet axial charge
and constrained by the $U(1)_A$ anomaly of QCD. This circumstance allows us to estimate
the nucleon matrix element of Eq.~(\ref{O6_def_G}) with minimal model dependence.
Our result is substantially larger than the estimate of Ref.~\cite{Bigi:1990kz},
because of the strong localization of the nonperturbative gluon fields in the instanton vacuum. 
We comment on the results of other approaches in light of our findings.

The object of the present study is the nucleon matrix element of the operator Eq.~(\ref{O6_def_G}),
not its role in CP-violation or the neutron EDM. Even so, our framework allows us to comment
on the neutron EDM induced by Eq.~(\ref{O6_def_G}). We argue that, in spite of the large
value of the nucleon matrix element of Eq.~(\ref{O6_def_G}), the instanton vacuum
suggests that the neutron EDM induced by the operator is suppressed and of similar
size as obtained in earlier estimates \cite{Bigi:1990kz}. These comments should be regarded
as speculative and inviting of further study.
\section{Calculation}
We consider the dimension-6 spin-0 CP-odd gluon operator as defined by
Eq.~(\ref{O6_def_G}). The field strength tensor is $(G_{\mu\nu})^a \equiv
\partial_\mu B^a_\nu - \partial_\nu B^a_\mu + g f^{abc} B^b_\mu B^c_\nu$, with the
gauge potential $B^a_\mu$ defined such that the covariant derivative for fermions is
$\nabla_\mu = \partial_\mu - i g B^a_\mu (\lambda^a/2)$. The dual field strength tensor is
$(\tilde G_{\mu\nu})^a \equiv (1/2) \epsilon_{\mu\nu\rho\sigma} (G^{\rho\sigma})^a$
with $\epsilon^{0123} = 1$ and Minkowskian metric $({+}{-}{-}{-})$.
As is standard in nonperturbative calculations, we use an alternative definition
of the field strength and potential, in which the coupling constant
is included in the fields, $F_{\mu\nu} \equiv g G_{\mu\nu}, A_\mu \equiv g B_\mu$,
and $\tilde F_{\mu\nu} \equiv g \tilde G_{\mu\nu}$;
these fields correspond to those in Ref.~\cite{Diakonov:1995qy} and the
instanton literature. In terms of these fields the operator
Eq.~(\ref{O6_def_G}) is identically expressed as
\begin{align}
\mathcal{O}_{\rm P6} = (1/16\pi^2) \,
f^{abc} (\tilde F_{\mu\nu})^a (F^{\mu\rho})^b (F^{\nu}_{\;\;\rho})^c .
\label{op_def}
\end{align}
The matrix element of the operator (at the space-time point $x = 0$) between nucleon
states with momenta $p$ and $p'$ is parameterized as
\begin{align}
\langle N (p') | \, \mathcal{O}_{\rm P6} \, | N(p) \rangle
&= A_{\rm P6}(q^2) \, m_N \bar U' i \gamma_5 U ,
\end{align}
where $q \equiv p' - p$ is the 4-momentum transfer, $m_N$ is the nucleon mass,
$U'$ and $U$ are the nucleon 4-spinors,
and $\gamma_5 \equiv -i \gamma^0 \gamma^1 \gamma^2 \gamma^3$ (sign opposite to the
Bjorken-Drell convention). The bilinear form vanishes at $q = 0$ and can be
represented as
\begin{align}
m_N \bar U' i \gamma_5 U \; = \; (i/2) \, S^\mu q_\mu,
\hspace{2em}
S^\mu \; \equiv \; \bar U' \gamma^\mu \gamma_5 U ,
\end{align}
where $S$ is the spin 4-vector of the $N\rightarrow N'$ transition.
The form factor $A_{\rm P6}$ is a function of the
invariant $q^2$ and has dimension $\textrm{(mass)}^2$. We prefer to
define it as a dimensionful quantity and not absorb the dimension
by powers of $m_N$, as $m_N$ is not the natural dynamical
scale for this matrix element (see below).

Parallel to the dimension-6 operator, Eqs.~(\ref{O6_def_G}) and (\ref{op_def}),
we consider the well-known dimension-4 operator
\begin{align}
\mathcal{O}_{\rm P4} \equiv
(g^2/16\pi^2) \, (\tilde G_{\mu\nu})^a (G^{\mu\nu})^a
\; = \;
(1/16\pi^2) \, (\tilde F_{\mu\nu})^a (F^{\mu\nu})^a ,
\label{O4_def}
\end{align}
whose matrix element is parameterized as
\begin{align}
\langle N (p') | \, \mathcal{O}_{\rm P4} \, | N(p) \rangle
&= A_{\rm P4}(q^2) \, m_N \bar U' i \gamma_5 U ,
\end{align}
where the form factor $A_{\rm P4}$ has dimension $\textrm{(mass)}^0$.
The matrix element can be inferred from the $U(1)_A$ anomaly in QCD. This operator relation
states that the divergence of the flavor-singlet axial current is given by
an anomalous term proportional to the gluonic operator Eq.~(\ref{O4_def}),
and a regular term proportional to the quark mass and pseudoscalar density,
\begin{align}
{\textstyle \sum_f} \partial_\mu (\bar\psi_f \gamma^\mu\gamma_5 \psi_f)
&= N_f \, \mathcal{O}_{\rm P4}
\; + \; 2 {\textstyle \sum_f} m_f \bar \psi_f i \gamma_5 \psi_f ,
\label{axial_anomaly}
\end{align}
where $\psi_f$ is the quark field, and the sum runs over the $N_f$ light quark flavors.
The nucleon matrix element of the flavor-singlet axial current is parameterized as
\begin{align}
&\langle N (p') | \, {\textstyle\sum_f} \bar\psi_f \gamma^\mu\gamma_5 \psi_f \, | N(p) \rangle
\nonumber \\
&= \bar U' \left[ \gamma^\mu\gamma_5 \, G_A^{(0)}(q^2)
- \frac{q^\mu \gamma_5}{2m_N} \, G_P^{(0)}(q^2) \right] U ,
\label{axial_me_q2}
\end{align}
where $G_A^{(0)}$ and $G_P^{(0)}$ are the axial and pseudoscalar form factors. Combining Eq.~(\ref{axial_anomaly})
and Eq.~(\ref{axial_me_q2}), neglecting the quark mass dependent term, one obtains
\begin{align}
A_{\rm P4}(q^2) &= \frac{2}{N_f} \left[ G_A^{(0)}(q^2) - \frac{q^2}{4 m_N^2} G_P^{(0)}(q^2) \right] .
\label{A4_from_axial_q2}
\end{align}
The flavor-singlet pseudoscalar form factor $G_P^{(0)}$ does not have a Goldstone boson pole,
so that the second term is suppressed at small $q^2$ even in the limit of zero light quark masses.
The form factor at $q^2 = 0$ is therefore given by (see e.g.\ Refs.~\cite{Bigi:1990kz,Diakonov:1995qy})
\begin{align}
A_{\rm P4}(0) \; &= \;
2 g_A^{(0)} / N_f ,
\label{A4_from_axial}
\end{align}
where $g_A^{(0)} \equiv G_A^{(0)}(0)$ is the nucleon flavor-singlet axial coupling. 

For evaluation in the instanton vacuum one expresses the operators Eqs.~(\ref{op_def})
and (\ref{O4_def}) in terms of the fields of the Euclidean theory, using the
standard conventions $(x^0, x^i) \rightarrow (\bar x_i, \bar x_4)$
with $x^0 = -i \bar x_4, x^i = \bar x_i$, $\bar x_\mu \bar x_\mu = \bar x_i^2 + \bar x_4^2$.
The Minkowskian fields are related to the Euclidean ones (denoted by a bar) as
\begin{align}
F_{0i} = i {\bar F}_{4i}, \; F_{ij} = {\bar F}_{ij}, \;
\tilde F_{0i} = \tilde {\bar F}_{4i}, \; \tilde F_{ij} = -i \tilde {\bar F}_{ij},
\end{align}
where $\tilde {\bar F}_{\mu\nu} \equiv (1/2) \epsilon_{\mu\nu\rho\sigma} {\bar F}_{\rho\sigma}$ with 
$\epsilon_{1234} = 1$. We obtain
\begin{align}
(-i) \, \mathcal{O}_{\rm P6} \; &= \; (1/16\pi^2) 
\, f^{abc} \tilde{\bar F}^a_{\mu\nu} \bar F^b_{\mu\rho} \bar F^c_{\nu\rho} ,
\label{O6_euclidean}
\\
(-i) \, \mathcal{O}_{\rm P4} \; &= \; (1/16\pi^2) 
\, \tilde {\bar F}^a_{\mu\nu} \bar F^a_{\mu\nu} .
\label{O4_euclidean}
\end{align}
In the following all coordinates and fields are understood as Euclidean, and the bar is omitted.

We now calculate the nucleon matrix elements of the operators in the instanton vacuum
following the approach of Ref.~\cite{Diakonov:1995qy}. The framework is the variational
approximation to the interacting instanton ensemble \cite{Diakonov:1983hh},
where all dynamical scales emerge
from the scale inherent in the QCD coupling, consistent with the renormalization properties of QCD.
The fermionic zero modes of the instantons cause the spontaneous breaking of chiral symmetry
and give rise to an effective theory of massive quarks with multifermionic
interactions \cite{Diakonov:1985eg}. Hadronic correlation functions are computed using the
$1/N_c$ expansion (saddle point approximation to the functional integrals), and the nucleon
emerges as a saddle point solution of massive quarks bound by a classical Goldstone boson
field (chiral soliton) \cite{Diakonov:1987ty}. Gluon operators in the original instanton
ensemble are systematically converted to ``effective operators'' in the effective theory
of massive quarks, where they can be inserted in hadronic correlation functions for calculation
of the matrix elements. The method relies on the smallness of the instanton packing fraction
$\bar\rho / \bar R$ and preserves the basic properties of the
QCD operators at the level of the effective operators. We outline the steps of the calculation
and refer to Ref.~\cite{Diakonov:1995qy} for a detailed description of the method.

In the first step, in leading order of $\bar\rho / \bar R$, the gluonic operators Eqs.~(\ref{O6_euclidean})
and (\ref{O4_euclidean}) are evaluated in the field of a single instanton or antiinstanton
($I$ or $\bar I$). The $I (\bar I)$ fields in singular gauge are
(with size $\rho$, standard color orientation, and center at $x = 0$)
\begin{align}
F^a_{\mu\nu}(x)_{I (\bar I)} \; =& \; 
(\eta^{\mp})^a_{\alpha\beta} \left(
  \frac{x_\mu x_\alpha}{x^2} \, \delta_{\nu\beta}
+ \frac{x_\nu x_\beta}{x^2} \, \delta_{\mu\alpha}
- \frac{1}{2} \delta_{\mu\alpha} \delta_{\nu\beta} \right)
\nonumber
\\ & \; \times \frac{8 \rho^2}{(x^2 + \rho^2)^2} ,
\label{F_inst}
\\
\tilde F^a_{\mu\nu}(x)_{I (\bar I)} \; =& \;
\pm F^a_{\mu\nu}(x)_{I (\bar I)} ,
\label{F_anti}
\end{align}
where $(\eta^\mp)^a_{\alpha\beta} \equiv \bar\eta^a_{\alpha\beta}, \eta^a_{\alpha\beta}$ are
the 't Hooft symbols. The $I (\bar I)$ fields take values in an $\textit{SU}(2)$ subalgebra
of the $\textit{SU}(N_c)$ color algebra, in which the structure constants are
$f^{abc} = \epsilon^{abc}$ (with $a,b,c$ in $\{1,2, 3\})$. Computing the color sum
in the dimension-6 operator Eq.~(\ref{O6_euclidean}), we obtain 
\begin{align}
f^{abc} {\tilde F}^a_{\mu\nu} F^b_{\mu\rho} F^c_{\nu\rho}(x)_{I ({\bar I})}
\; &= \; \mp \frac{3 \cdot 512 \, \rho^6}{(x^2 + \rho^2)^6}
\; \equiv \; \mp f_6(x \, | \, \rho).
\label{O6_inst}
\end{align}
This should be compared with the well-known result for the dimension-4 operator Eq.~(\ref{O4_euclidean})
in the $I (\bar I)$ field,\footnote{Because the gluon operators in Eqs.~(\ref{O6_inst}) and (\ref{O4_inst})
are gauge-invariant and do not explicitly involve fermions,
they can as well be evaluated using the instanton field in regular gauge,
which has a simpler form than Eq.~(\ref{F_inst}).}
\begin{align}
{\tilde F}^a_{\mu\nu} F^a_{\mu\nu}(x)_{I ({\bar I})}
\; &= \; \pm \frac{3 \cdot 64 \, \rho^4}{(x^2 + \rho^2)^4}
\; \equiv \; \pm f_4(x \, | \, \rho),
\label{O4_inst}
\end{align}
which, up to a factor, is the $I (\bar I)$'s topological charge density.

In the second step, the effective fermion operators are constructed, by mutiplying
the gluonic operators in the $I (\bar I)$ field with the projectors on the fermionic zero modes,
and averaging over the collective coordinates of the $I (\bar I)$
(size, color orientation, center coordinate) \cite{Diakonov:1995qy}.
The average over the $I (\bar I)$ size is computed with the variational size 
distribution of Refs.~\cite{Diakonov:1983hh,Diakonov:1995qy}, which describes
the suppression of large $\rho$ by the instanton interactions in the ensemble;
small $\rho$ are suppressed by the vanishing free instanton weight.\footnote{The dimension-6 operator in the
instanton field, Eq.~(\ref{O6_inst}), scales as $f_6 (x \, | \, \rho) \sim \rho^{-6}$
when evaluated at $x \sim \rho$, inside the instanton. The instanton size distribution at small $\rho$
behaves as $d_\pm(\rho) \sim \rho^{b - 5} \log (\Lambda \rho)$, where $b = (11/3) \, N_c$
is the coefficient of the $\beta$ function in gluodynamics. Because $b \sim N_c$ is
parametrically large in the approach of Refs.~\cite{Diakonov:1983hh,Diakonov:1995qy},
the integral over $\rho$ converges at small $\rho$,
and the contributions from $\rho \ll \bar\rho$ are suppressed, even
when averaging higher-dimensional operators such as Eq.~(\ref{O6_inst}). It is obvious that
contributions from instantons with $\rho \ll \bar\rho$ should be suppressed if the QCD operator
is regarded as normalized at the scale $\mu = \bar\rho^{-1}$.}
The distribution is concentrated around $\rho \approx \bar\rho$,
and its width is parametrically small, so that the average is computed simply by replacing the
actual size $\rho$ by the average size $\bar\rho$ in the operators.
Performing the average over the color orientation and the center coordinate $z$,
we obtain the effective operators corresponding to $\mathcal{O}_{\rm P6}$ and
$\mathcal{O}_{\rm P4}$ (denoted by quotation marks) as
\begin{align}
\textrm{``$\mathcal{O}_{\rm P6}(x)$"} \; =& \;
\frac{\lambda}{16 \pi^2}
\int d^4 z \; f_6 (x - z \, | \, \bar\rho)
\nonumber 
\\
& \times
\left[ -\textrm{det} \, iJ_+(z) + \textrm{det} \, iJ_-(z) \right] ,
\label{O6_eff}
\\
\textrm{``$\mathcal{O}_{\rm P4}(x)$"} \; =& \;
\frac{\lambda}{16 \pi^2}
\int d^4 z \; f_4 (x - z \, | \, \bar\rho)
\nonumber 
\\
& \times
\left[ \textrm{det} \, iJ_+(z) - \textrm{det} \, iJ_-(z) \right] ,
\label{O4_eff}
\\
\lambda \; \equiv& \; (2V/N)^{N_f - 1} M^{N_f} .
\label{lambda}
\end{align}
Here $\lambda$ is a dimensionful factor composed from the average instanton density $N/(2V)$ and the
dynamical quark mass generated by chiral symmetry breaking, $M$; the factor represents the dynamical
scales emerging from the instanton vacuum and provides the normalization of the
effective operator \cite{Diakonov:1995qy}. (The present calculation is performed with an equal number of
$I$ and $\bar I$ in the ensemble, $N_+ = N_- = N/2$, which is sufficient for connected
correlation functions of the operators at nonzero momentum transfer; the role of topological
fluctuations $N_+ \neq N_-$ is discussed below.) The functions
\begin{align}
J_\pm (z)_{fg} &= {\psi^\dagger}_{\!\!\! f} (z) \, F({\stackrel{\leftarrow}{{}\partial}})
{\textstyle\frac{1}{2}} (1 \pm \gamma_5)
F({\stackrel{\rightarrow}{{}\partial}}) \,
\psi_g (z)
\label{J_flavor}
\end{align}
are the left- and right-handed chiral quark densities, where
$\psi^\dagger \equiv i \bar\psi$ and $\psi$ are the Euclidean quark fields, and $F(...)$ denote the form factors
resulting from the $I (\bar I)$ fermionic zero modes of the $I(\bar I)$ \cite{Diakonov:1995qy}.
The $J_\pm (z)$ are $N_f \times N_f$ matrices
in flavor, and the fermionic vertices appearing in the effective operators
Eqs.~(\ref{O6_eff}) and (\ref{O4_eff}) have the same flavor determinant form as the
't Hooft vertices appearing in the effective fermion action derived from instantons \cite{Diakonov:1995qy}.
The differences between the left- and right-handed vertices appear because
the gluonic operators are CP-odd and take opposite values in the $I$ and $\bar I$.

In the final step, the effective operators are inserted in hadronic correlation functions
evaluated in the effective theory of massive quarks. If the momentum transfer to the
gluonic operator is of the order $q \sim \bar R^{-1} \ll \bar \rho^{-1}$, the non-locality of
the instanton field in Eqs.~(\ref{O6_eff}) and (\ref{O4_eff}) can be neglected,
and we can approximate the profile functions $f_6$ and $f_4$ by 4-dimensional
$\delta$ functions, with coefficients fixed by the integral over $x$,
\begin{align}
f_{6, 4} (x - z) \rightarrow C_{6, 4} \delta^{(4)} (x - z),
\hspace{1em}
C_{6, 4} \equiv \int d^4 x \; f_{6, 4} (x) ; 
\end{align}
the difference between the infinite integral and a finite integral
over a domain $x \sim \bar R$ is a correction in $\bar\rho/\bar R$ because of the localization
of the instanton field. Integrating the profile functions Eqs.~(\ref{O6_inst})
and (\ref{O4_inst}) over $x$, we obtain
\begin{align}
C_6 \; = \; 3 \cdot 128 \pi^2 / (5 \, \bar\rho^2), \hspace{2em}
C_4 \; = \; 32 \pi^2 ;
\label{O4_int_inst}
\end{align}
the value of $C_4$ is the well-known instanton action.
In this approximation the effective operators become
\begin{align}
\left.
\!\!\!
\begin{array}{l}
\textrm{``$\mathcal{O}_{\rm P6}(x)$"}
\\[1ex]
\textrm{``$\mathcal{O}_{\rm P4}(x)$"}
\end{array} \!
\right\}
=&
\left\{ \!
\begin{array}{r}
-C_6
\\[1ex]
C_4
\end{array} \!
\right\}
\times \frac{\lambda}{16\pi^2} \left[ \textrm{det} \, i J_+(x) - \textrm{det} \, i J_-(x) \right] ,
\label{O6_O4_eff_local}
\end{align}
and one sees that the effective operators for $\mathcal{O}_{\rm P6}$ and $\mathcal{O}_{\rm P4}$ are
the same up to the coefficients. Without calculating the nucleon correlation function of the operators,
we therefore immediately conclude that the nucleon matrix elements are related by
\begin{align}
\frac{A_{\rm P6}(q^2)}{A_{\rm P4}(q^2)} \; = \; - \frac{C_6}{C_4}
\; = \; - \frac{12}{5 \bar\rho^2}
\hspace{2em} (q \sim \bar R^{-1}).
\label{O6_O4_relation}
\end{align}
This conclusion relies only on the fact that the operators enter in the correlation
functions by coupling to a single $I (\bar I)$, and on the specific value of the operators in the
$I (\bar I)$ fields. Now combining Eq.~(\ref{O6_O4_relation}) with the value of $A_{\rm P4}(0)$
obtained from the $U(1)_A$ anomaly, Eq.~(\ref{axial_anomaly}), we obtain
\begin{align}
A_{\rm P6}(0) &= - \frac{12}{5 \bar\rho^2} \times \frac{2 g_A^{(0)}}{N_f} .
\label{O6_result}
\end{align}
This estimate uses the instanton vacuum only to relate the dimension-6 and 4 operators at the
instanton scale but does not require the calculation of $g_A^{(0)}$ in the instanton vacuum.

Alternatively, we could estimate $A_{\rm P6}(0)$ by performing the actual calculation of
the nucleon matrix element of the effective operators Eq.~(\ref{O6_O4_eff_local}) in the
instanton vacuum. The result would be identical to Eq.~(\ref{O6_result}), only with $g_A^{(0)}$
replaced by the value obtained from the instanton vacuum. This happens because
the nucleon matrix element of the gluonic operator $\mathcal{O}_{\rm P4}$ calculated in the instanton vacuum
is proportional to that of the divergence of the flavor-singlet axial current in the instanton vacuum,
leading to the same relation as obtained from the $U(1)_A$ anomaly in QCD \cite{Diakonov:1995qy}.
This remarkable result comes about as follows. The effective action of massive quarks
emerging from chiral symmetry breaking in the instanton vacuum has the form
\begin{align}
-S_{\rm eff} &= \int d^4 x \left\{ \psi^\dagger i \partial\gamma \psi (x)
+ \lambda \left[ \textrm{det} \, i J_+(x) + \textrm{det} \, i J_-(x) \right] \right\} ,
\label{S_eff}
\end{align}
where $\lambda$ is the dynamical scale Eq.~(\ref{lambda}).
The flavor-singlet axial current in the effective theory and its divergence can be
derived from the response to a space-time dependent $U(1)_A$ chiral transformation
of the quark fields,
\begin{align}
\psi, \, \psi^\dagger \; \rightarrow \; \exp(i\epsilon \gamma_5) \, \psi, \;
\psi^\dagger \exp(i\epsilon \gamma_5), \hspace{1em} \epsilon \equiv \epsilon(x),
\end{align}
and one obtains
\begin{align}
J_{5\mu}(x)_{\rm eff} &= {\textstyle\sum_f} \, {\psi^\dagger}_{\!\!\! f} \gamma_\mu\gamma_5 \psi_f (x),
\\
\partial_\mu J_{5\mu}(x)_{\rm eff}
&= 2 N_f  \; \lambda \left[ \textrm{det} \, i J_+(x) - \textrm{det} \, i J_-(x) \right] .
\end{align}
The divergence has the same form as the effective operator for $\mathcal{O}_4$ in the
effective theory, Eq.~(\ref{O6_O4_eff_local}), with $C_4$ given by Eq.~(\ref{O4_int_inst}).
The instanton vacuum thus provides
the same relation between the flavor-singlet axial current and the gluonic operator $\mathcal{O}_4$
as the $U(1)_A$ anomaly in QCD, Eq.~(\ref{axial_anomaly}). Note, however, that the nature of $U(1)_A$
symmetry breaking is is very different in the two cases. In QCD the $U(1)_A$ symmetry is broken
anomalously, by the integration over high-momentum modes of the fermion fields.
The effective theory derived from instantons contains only low-momentum modes
(momenta $k \lesssim \bar\rho^{-1}$) and has no anomaly, but the $U(1)_A$ symmetry
is broken explicitly by the dynamical scales embedded in the
theory.\footnote{It is worth emphasizing that the $U(1)_A$ anomaly in QCD involves only
the dimension-4 gluon operator $\mathcal{O}_{\rm P4}$, not any higher-dimensional operators.
The relation between $\mathcal{O}_{\rm P6}$ and $\mathcal{O}_{\rm P4}$ observed in the instanton vacuum,
Eq.~(\ref{O6_O4_relation}), is specific to the effective theory with cutoff $\bar\rho^{-2}$ and
does not imply that $\mathcal{O}_{\rm P6}$ would somehow participate in the QCD anomaly
along with $\mathcal{O}_{\rm P4}$.}

We can use Eq.~(\ref{O6_result}) to estimate the numerical
value of $A_{\rm P6}(0)$. The ratio between the dimension-6 and 4 matrix elements is
(using $\bar\rho^{-1} = 0.60$ GeV)
\begin{align}
12/(5 \bar\rho^2)
\; = \; 0.86 \, \textrm{GeV}^2 \; = \; (0.22 \, \textrm{fm})^{-2} .
\label{ratio_num}
\end{align}
This represents a large value on the hadronic scale and attests to the strength of the
localized non-perturbative gluon fields in the instanton vacuum. The flavor-singlet
axial coupling $g_A^{(0)}$ can be determined in polarized DIS
experiments (where it appears as the total light quark contribution
to the nucleon spin, $\Delta\Sigma$) or in theoretical calculations. The polarized DIS analysis of
Ref.~\cite{Ethier:2017zbq} with three light flavors ($N_f = 3$) obtains $g_A^{(0)} = 0.36(9)$
at the scale $\mu = 1$ GeV; other analyses report values in the range 0.20--0.25
with uncertainties up to $\sim$100\%; see Ref.~\cite{Ball:2013lla} for a compilation.
Recent Lattice QCD calculations obtain values of $g_A^{(0)}$ in the range
$\sim$0.3--0.45 \cite{Liang:2018pis,Alexandrou:2019brg,Alexandrou:2020sml}.
A calculation in the chiral soliton model of the nucleon, based on the effective action Eq.~(\ref{S_eff}),
gives $g_A^{(0)} = 0.36$ at the scale $\mu \sim \bar\rho^{-1}$ \cite{Blotz:1993am}.
For a conservative estimate covering these values we assume
\begin{align}
g_A^{(0)} = 0.36^{+0.09}_{-0.36} \hspace{2em} (N_f = 3)
\end{align}
and neglect the scale dependence compared to the uncertainties. With these parameters
we obtain
\begin{align}
A_{\rm P6}(0) &= - (0.21^{+0.05}_{-0.21}) \, \textrm{GeV}^2 .
\label{O6_numerical}
\end{align}
This estimate refers to the instanton vacuum scale $\mu \sim \bar\rho^{-1}$.

To justify our instanton vacuum calculation of the $\mathcal{O}_{\rm P6}$ and $\mathcal{O}_{\rm P4}$
matrix elements, we need to comment on the role of topological charge fluctuations.
In Ref.~\cite{Diakonov:1995qy} the instanton vacuum
was constructed as a grand canonical ensemble with fluctuating numbers of $I$'s and $\bar I$'s,
including topological charge fluctuations with $\Delta \equiv N_+ - N_- \neq 0$.
It was shown that the topological susceptibility $\langle \Delta^2 \rangle$ is qualitatively
affected by the fermion determinant and vanishes if at least one quark
flavor becomes massless, in accordance with general expectations. The effective operator
formalism was developed including $\Delta$ fluctuations, which are essential for correlation
functions of topological operators at zero momentum transfer (integrated over the system volume $V$),
such as the topological charge $\int_V d^4 x \, \tilde FF (x)$. In the present study we
consider the connected correlation functions of the local operator $\tilde F F (x)$
at non-zero momentum transfer, which are not affected by $\Delta$ fluctuations and
can be computed in the effective theory at $\Delta = 0$; the fluctuations experienced by the
operator in a finite subvolume do not depend on the boundary conditions of the system \cite{Diakonov:1995qy}.
In this sense the present calculation is consistent with 
the general treatment of Ref.~\cite{Diakonov:1995qy}. Note, however, that when one considers
correlation functions of $\mathcal{O}_{\rm P6}$ and $\mathcal{O}_{\rm P4}$
integrated over the system volume, one has to revert back to the
ensemble with $\Delta$ fluctuations \cite{Faccioli:2004jz}.

\section{Discussion}
It is interesting to compare our estimate with those of other approaches.
In Ref.~\cite{Bigi:1990kz} $A_{\rm P6}(0)$ was estimated by assuming that the ratio of the
nucleon matrix elements of the dimension-6 and dimension-4 pseudoscalar gluon operators
is the same as the ratio of the vacuum matrix elements of the corresponding scalar gluon operators;
in short-hand notation:
\begin{align}
\frac{A_{\rm P6}(0)}{A_{\rm P4}(0)}
= \frac{\langle N| f {\tilde F} F F| N \rangle}{\langle N| {\tilde F} F| N \rangle}
\approx \frac{\langle 0| f F F F| 0 \rangle}{\langle 0| F F| 0 \rangle} .
\label{ratio_bigi}
\end{align}
The vacuum ratio was then evaluated using the empirical values of the gluon condensates
determined in QCD sum rule calculations, resulting in $A_{\rm P6}(0)/A_{\rm P4}(0) \approx 0.13$ GeV$^2$.
In our instanton-based approach Eq.~(\ref{ratio_bigi}) is indeed realized
in leading order of the packing fraction, if the vacuum condensates are saturated by the
instanton fields; this comes about because the $I (\bar I)$ fields are self-dual
(anti-self-dual), cf.\ Eq.~(\ref{F_anti}).
However, the numerical value of the ratio is obtained as in Eqs.~(\ref{O6_O4_relation})
and (\ref{ratio_num}), $\sim$7 times larger than the estimate
of Ref.~\cite{Bigi:1990kz}. The enhancement is due to the strong localization of the non-perturbative
gluon fields in the instanton vacuum, which increases the value of higher-dimensional condensates
relative to expectations based on a uniform distribution.

In Refs.~\cite{Hatta:2020ltd,Hatta:2020riw} the gluon operator $\mathcal{O}_{\rm P6}$ was related to
other dimension-6 operators using the operator identity [expressed here in our conventions
and for Minkowskian fields, cf.\ Eq.~(\ref{op_def})]
\begin{align}
& \; (1/16\pi^2) \, {\textstyle \sum_f}
\partial^\mu \left[ \bar\psi_f (\tilde F_{\mu\nu})^a \, (\lambda^a/2) \, \gamma^\nu \psi_f
\right]
\nonumber
\\
=& \; (1/16\pi^2) \,
f^{abc} (\tilde F_{\mu\nu})^a (F^{\mu\rho})^b (F^{\nu}_{\;\;\rho})^c
\nonumber
\\
-& \; (1/32\pi^2) \, 
({\tilde F}_{\mu\nu})^a (D_\rho)^{ab} (D^\rho)^{bc} (F^{\mu\nu})^c
\nonumber
\\[1ex]
\equiv& \; \mathcal{O}_{\rm P6} + \mathcal{O}_{\rm D} ,
\label{operator_relation}
\end{align}
where $(D_\rho)^{ab} \equiv \delta^{ab} \partial_\rho  + f^{aib} A_\rho^i$ is the covariant derivative
in the adjoint representation. Equation~(\ref{operator_relation}) is obtained from the QCD equations
of motion for the fields and the algebraic identities for covariant derivatives.
The dimension-5 spin-1 quark-gluon operator (the operator under the total derivative) appears in the
twist-4 corrections to the isoscalar nucleon spin structure functions \cite{Shuryak:1981pi},
and its matrix element can
be inferred from DIS data and theoretical calculations. The matrix element of $\mathcal{O}_{\rm P6}$
was then estimated in Ref.~\cite{Hatta:2020riw} assuming that the contributions of all three operators
are of the same magnitude at the hadronic scale.
The instanton vacuum suggests that this assumption is not justified, for the
following reasons: (i)~The matrix element of $\mathcal{O}_{\rm P6}$ estimated
in Ref.~\cite{Hatta:2020riw} is substantially smaller than our instanton vacuum result.
Expressed in our convention, the estimate of Ref.~\cite{Hatta:2020riw} is
\begin{align}
A_{\rm P6}(0) = - (g^2/16\pi^2) \, 4 m_N^2 E(0) \approx - (0.18 \, \textrm{GeV}^2) \times E(0) ,
\end{align}
where $E$ is the form factor introduced in Ref.~\cite{Hatta:2020riw};
in the last equation we have evaluated the QCD coupling $g^2$ using the well-known LO expression
($N_f = 3, \Lambda =$ 0.2 GeV, $\mu = \bar\rho^{-1}$ = 0.6 GeV). With $E(0) \sim 0.01$ as used
in Ref.~\cite{Hatta:2020riw}, $A_{\rm P6}(0)$ would be two orders of magnitude smaller than
our result Eq.~(\ref{O6_numerical}). (ii)~The instanton vacuum predicts that the nucleon matrix elements
of $\mathcal{O}_{\rm P6}$ and $\mathcal{O}_{\rm D}$ are individually
large but cancel each other in Eq.~(\ref{operator_relation}). Extending our calculation of the
matrix element from $\mathcal{O}_{\rm P6}$ to $\mathcal{O}_{\rm D}$, we obtain
\begin{align}
A_{\rm D}(0) = -A_{\rm P6}(0)
\hspace{2em} (\textrm{in leading order of $\bar\rho/\bar R$}).
\end{align}
In fact, the cancellation of $\mathcal{O}_{\rm P6}$ and $\mathcal{O}_{\rm D}$ in one-instanton
approximation is mathematically necessary, because the instanton is a (complex) solution to the
free Yang-Mills equations without quark sources, and using the free equations in deriving the
operator identity one would obtain zero instead of the quark-gluon operator in Eq.~(\ref{operator_relation}).
We note that the cancellation of $\mathcal{O}_{\rm P6}$ and $\mathcal{O}_{\rm D}$ is consistent with the
earlier instanton vacuum result for the nucleon matrix element of the flavor-singlet twist-4
quark-gluon operator, which is of the order $M^2 \sim \bar\rho^2/\bar R^4$ rather than $\bar\rho^{-2}$ and thus
parametrically suppressed \cite{Lee:2001ug}. (In contrast, the flavor-nonsinglet nucleon matrix
element is of order $\bar\rho^{-2}$ and parametrically large \cite{Balla:1997hf}.)
In summary, the instanton vacuum shows that the matrix elements of the operators in
Eq.~(\ref{operator_relation}) have very unequal magnitude, and that one cannot infer the
``large'' gluonic ones from the ``small'' quark-gluon one.

We also want to comment on the implications of our results for CP-violation and the neutron EDM.
The scenario of Ref.~\cite{Weinberg:1989dx} considers a CP-violating Lagrangian of the form
\begin{align}
\delta L_{\rm CP} \; = \; a_6 \mathcal{O}_{\rm P6} , 
\label{CP_6}
\end{align}
where the coefficient $a_6$ results from the CP-violating short-distance processes and
includes the effects of renormalization group evolution to the hadronic scale. The nucleon
EDM appears through the correlation function
\begin{align}
i \int d^4 x \, 
\langle N' | \, \textrm{T} \, \mathcal{O}_{\rm P6} (x) \, J^\mu_{\rm em} (0) \, | N \rangle ,
\label{EDM}
\end{align}
which describes the electromagnetic vertex of the nucleon under the influence of the
CP-odd Lagrangian. The correlation function can be computed by inserting a complete
set of hadronic intermediate states, which include the nucleon state and ``other'' states
(whose precise composition is not relevant),
\begin{align}
\sum_X | X \rangle \langle X| \; &= \; | N \rangle \langle N| \; + \; \sum_{X' \neq N} | X' \rangle \langle X'| .
\label{states}
\end{align}
In Ref.~\cite{Bigi:1990kz} it was assumed that Eq.~(\ref{EDM}) can be approximated by the
contribution of the nucleon intermediate state, i.e., that there are no strong cancellations
between the nucleon and the other intermediate states. This made it possible to compute
the correlation function in in terms of the nucleon matrix elements
$\langle N' | \mathcal{O}_{\rm P6} | N \rangle$ and $\langle N' | J^\mu_{\rm em} | N \rangle$;
see Ref.~\cite{Hatta:2020riw} for details. The instanton vacuum suggests
that this assumption is likely not valid. When Eq.~(\ref{EDM}) is computed in the
instanton vacuum, we can suppose that the effective operator representing $\mathcal{O}_{\rm P6}$
is proportional to that of $\mathcal{O}_{\rm P4}$, see Eq.~(\ref{O6_O4_eff_local}). The EDM
extracted from the correlation function is thus proportional to that induced by
$\mathcal{O}_{\rm P4}$, the traditional $\theta$ angle term. The latter is known
to be proportional to the quark mass and suppressed in the chiral limit \cite{Crewther:1979pi}.
This chiral suppression requires the exact cancellation of the contributions of the nucleon and
other intermediate states in Eq.~(\ref{EDM}) and is incompatible with the intermediate-nucleon
approximation, as was already pointed out in Ref.~\cite{Bigi:1990kz}.\footnote{In the instanton
vacuum the correlation function Eq.~(\ref{EDM}) is of course computed without projection on hadronic
intermediate states, using the effective theory of massive quarks. The point here is that the
properties of the effective operators suggest that the EDM resulting from $\mathcal{O}_{\rm P6}$
is chirally suppressed. This circumstance then implies that, in the hadronic representation,
there are cancellations between the nucleon and other intermediate states.}
One therefore should not use our result for the nucleon matrix element of
$\mathcal{O}_{\rm P6}$, Eq.~(\ref{O6_numerical}),
to estimate the EDM in the approach of Refs.~\cite{Bigi:1990kz,Hatta:2020riw};
doing so would grossly overestimate it.

Instead, our findings suggest a different way to estimate the neutron EDM induced by
$\mathcal{O}_{\rm P6}$. Assuming that the proportionality between the effective operators
for $\mathcal{O}_{\rm P6}$ and $\mathcal{O}_{\rm P4}$ holds when computing the full correlation
function Eq.~(\ref{EDM}) in the instanton vacuum, we can estimate the EDM induced by
$\mathcal{O}_{\rm P6}$ by ``converting'' previous results for the EDM induced by $\mathcal{O}_{\rm P4}$.
The CP-violating Lagrangian associated with $\mathcal{O}_{\rm P4}$ is
\begin{align}
\delta L_{\rm CP} \; = \; (\theta/2) \, \mathcal{O}_{\rm P4} ,
\label{CP_4}
\end{align}
where $\theta$ is the vacuum angle. Choosing $\theta/2 =  (-12/5\bar\rho^2)\, a_6$,
we effectively match the strength of Eq.~(\ref{CP_4}) to Eq.~(\ref{CP_6}) in the
instanton vacuum correlation functions; see Eqs.(\ref{O6_O4_eff_local}) and
(\ref{O6_O4_relation}).\footnote{It goes without saying that this identification
makes sense only in the context of the effective operators in the instanton vacuum
and does not imply a general relation between the QCD operators.}
For a numerical estimate of the neutron EDM induced by Eq.~(\ref{CP_4}) we use the well-known result of
Ref.~\cite{Crewther:1979pi} based on the non-analytic chiral term,
$|d_n| = 3.6 \times 10^{-3}\; |\theta| \; e \; \textrm{fm}$,
which takes into account the chiral suppression of the EDM. In this way we estimate the neutron EDM
induced by Eq.~(\ref{CP_6}) in the instanton vacuum as
\begin{align}
|d_n| \; \sim \; 6 \times 10^{-3} \; |a_6 \cdot \textrm{GeV}^2| \; e \; \textrm{fm} .
\label{d_n}
\end{align}
This estimate is 4 times smaller than that of Ref.~\cite{Bigi:1990kz} obtained with the
intermediate-nucleon approximation, even though our nucleon matrix element of
$\mathcal{O}_{\rm P6}$ is larger by almost an order of magnitude (see above). While we do not
propose Eq.~(\ref{d_n}) as a serious estimate of the neutron EDM, it illustrates the limits of
the intermediate-nucleon approximation and demonstrates that the large nucleon
matrix element of $\mathcal{O}_{\rm P6}$ obtained from instantons is compatible with a normal
size of the neutron EDM.

A quantitative estimate of the EDM induced by $\mathcal{O}_{\rm P6}$ should be performed by
calculating the full correlation function Eq.~(\ref{EDM}) in the instanton vacuum and studying
its quark mass dependence. Such a calculation should confirm the chiral suppression of the
$\mathcal{O}_{\rm P6}$ EDM conjectured here and determine both analytic and non-analytic
terms in the quark mass dependence.

In summary, the instanton vacuum suggests an interesting picture of hadronic CP-violation induced
by the dimension-6 operator $\mathcal{O}_{\rm P6}$. The nonperturbative dynamics effectively
connects the operator with the topological charge density $\mathcal{O}_{\rm P4}$. The nucleon matrix
element of $\mathcal{O}_{\rm P6}$ is large because of the strong localization of the instanton field. 
However, the neutron EDM induced by $\mathcal{O}_{\rm P6}$ is of conventional size, because it is
subject to the same chiral suppression as that induced by $\mathcal{O}_{\rm P4}$. The picture
should be explored in further studies.

I wish to thank Y.~Hatta for exchanges initiating this study and greatly helping its course;
M.~V.~Polyakov, for valuable comments; and K.~Orginos, D.~G.~Richards, and N.~Sato,
for helpful communication. This material is based upon work supported by the U.S.~Department of Energy, 
Office of Science, Office of Nuclear Physics under contract DE-AC05-06OR23177.
\section*{References}
\end{document}